\documentclass[journal,onecolumn]{IEEEtran}
\usepackage[cmex10]{amsmath}
\usepackage{amssymb}
\usepackage[official]{eurosym}
\usepackage[dvips]{graphicx}
\usepackage{color}

\begin{document}

\title{Invariant EKF Design for Scan Matching-aided Localization}

\author{Martin~Barczyk,~\IEEEmembership{Member,~IEEE,}
        Silv{\`e}re~Bonnabel,
        Jean-Emmanuel~Deschaud,
        and~Fran{\c c}ois~Goulette,~\IEEEmembership{Member,~IEEE}
\thanks{M. Barczyk is with the Department of Mechanical Engineering, University of Alberta, Edmonton, AB, T6G 2G8 Canada \texttt{martin.barczyk@ualberta.ca}.}%
        \thanks{S. Bonnabel, J-E. Deschaud and F. Goulette are with MINES ParisTech, PSL - Research University, Centre de robotique, 60 Bd St Michel 75006 Paris, France \texttt{\{firstname.lastname\}@mines-paristech.fr}}}

\markboth{IEEE Transactions on Control Systems Technology}%
{Barczyk \MakeLowercase{\textit{et al.}}: Invariant EKF Design for Scan Matching-aided Localization}

\maketitle

\begin{abstract}
Localization in indoor environments is a technique which estimates the robot's pose by fusing data from onboard motion sensors with readings of the 
environment, in our case obtained by scan matching point clouds captured by a low-cost Kinect depth camera. We develop
both
an Invariant Extended Kalman Filter (IEKF)-based and a Multiplicative Extended Kalman Filter (MEKF)-based solution to this problem. The two
designs are successfully validated in experiments and demonstrate the advantage of the IEKF design.
\end{abstract}


\section{Introduction and Literature Review}
\label{sec:intro}

\IEEEPARstart{L}{ocalization} is a fundamental building block for mobile robotics~\cite{TBF05}. For operations in environments where GPS is not available 
e.g.~indoor navigation, or for situations where GPS is available but degraded, the uncertainty in the vehicle pose can be reduced by using information from 
a 
3D map~\cite{ZF13}. Fundamentally this represents
a sensor fusion problem,
and as such
it is typically handled via an Extended Kalman Filter (EKF) c.f.~the textbooks~\cite{Brown,Farrell}, although a number of direct nonlinear observer designs have 
been proposed e.g.~\cite{CHMB07,VCSO10,SMS11}. The EKF approach offers two practical advantages over these: it is a fully systematic design procedure, and 
it admits a probabilistic interpretation. Indeed it can be viewed as a maximum likelihood estimator that combines sensor information in an optimal way (in the 
linear case), with the confidence in each measurement being described by a covariance matrix.

In this paper, we address a localization problem, where pose estimates from proprioceptive sensors are obtained by matching images obtained from a depth 
scanner (scan matching or LIDAR-based odometry~\cite{LM94}) with a 3D map. We use the Iterative Closest-Point (ICP) algorithm~\cite{BM92,CM92} on 3D point 
clouds obtained from a low-cost Kinect camera mounted on our experimental wheeled robot. During the motion we assume an already built 3D gross map made of 
clouds of points is available. When a scan contains substantially more information than the map, it can be aggregated to the existing map. Further details will 
be provided in Section~\ref{sec:visionaiding}. 

The sensor fusion EKF works by linearizing a system about its estimated trajectory, then using estimates obtained from an observer for this linearized model
to correct the state
of the original system. In this way the EKF relies on a closed loop which can be destabilized by sufficiently poor estimates of the trajectory, known
as divergence~\cite{Brown}. Clearly, reducing or eliminating the dependence of the EKF on the system's trajectory would increase the
robustness of the
overall system. An emerging methodology to accomplish this goal is the Invariant EKF~\cite{Bon07,BMS09}, built on the theoretical foundations of invariant
(symmetry-preserving) observers~\cite{BMR08,BMR09}. The IEKF technique has already
demonstrated experimental performance improvements over a typical~\cite{Farrell} ``Multiplicative''
EKF (MEKF) in aided inertial navigation designs~\cite{MS10,BL13}. From a practical viewpoint, the advantage of EKF-like techniques is their automatic 
tuning of gains based on the estimated covariance of the measurements. Applying the IEKF to an aided scan matching problem was first 
demonstrated in~\cite{HBG12} for 3D 
mapping. A preliminary version of this work appears in conference proceedings~\cite{BBDG14}. The contributions of the present paper are as follows:
\begin{enumerate}
\item Designing a scan matching-aided localization system for a wheeled indoor robot which fuses robot odometry with Kinect-based scan matching,
\item A method to compute a realistic covariance matrix associated to the scan matching step and implicitly detect underconstrained 
environments,
\item Providing both an IEKF (whose non-linear structure takes advantage of the the geometry of the problem) and a MEKF version of the state estimator for this 
problem,
\item Experimentally validating both designs, and comparing their accuracy w.r.t.~ground truth data provided by an optical motion capture
system.
\end{enumerate}

This paper is structured as follows. A brief literature review and problem motivation have been provided above. Section~\ref{sec:systemmodels}
provides the model of
the system dynamics and derives linearizing approximations used in the sequel. Section~\ref{sec:visionaiding} covers the process of obtaining aiding 
measurements
from scan matching of point clouds provided by a Kinect depth camera mounted on the robot, and provides a meaningful and simple method to compute a covariance 
matrix associated to the measurements. Section~\ref{sec:estimatordesign} provides the equations of the
invariant observer, IEKF and MEKF state estimators, which are then experimentally validated in Section~\ref{sec:expvalidation}. Conclusions and future work
directions are given in Section~\ref{sec:conclusions}.

\section{System Models}
\label{sec:systemmodels}

\subsection{Noiseless System}
\label{sec:syseqs}

In the absence of sensor noise, the dynamics of a rigid-body vehicle are governed by
\begin{equation}
\begin{aligned}
\dot{R}&=R S(\omega)\\
\dot{p}&= R \mu
\end{aligned}
\label{eq:dyncomponents}
\end{equation}
where $R\in SO(3)$ is a rotation matrix measuring the 3D attitude of the vehicle, $\omega
\in
\mathbb{R}^3$ is the body-frame angular
velocity vector, $S(\cdot)$ is the $3\times 3$ skew-symmetric matrix such that $S(x)y=x \times y$
where
$\times$ denotes the $\mathbb{R}^3$ cross-product, $p\in\mathbb{R}^3$ is the position vector of the vehicle expressed in coordinates of the ground-fixed frame,
and $\mu\in\mathbb{R}^3$ is the velocity vector of the vehicle expressed in body-frame coordinates. The body-frame angular velocity $\omega$ and linear
velocity $\mu$ vectors can be measured directly using on-board motion sensors, e.g.~a triaxial rate gyro for $\omega$ and a Doppler radar for $\mu$. In the case
of a
wheeled vehicle traveling over flat terrain, odometry information from the left and right wheels can be used to compute the forward component of vector $\mu$
and the vertical component of vector $\omega$, taking the remaining components as zero.

As will be discussed in Section~\ref{sec:visionaiding} the vehicle's attitude $R$ and position $p$ are computed by \emph{scan
matching} images from the on-board Kinect depth camera with a map of 3D points via the Iterative Closest Point (ICP) algorithm. This gives output equations
\begin{equation}
\begin{pmatrix} y_R \\ y_p \end{pmatrix}
= \begin{pmatrix} R \\ p \end{pmatrix}
\label{eq:outcomponents}
\end{equation}

\subsection{Sensor noise models}
\label{sec:noise}

In reality the input signals $\omega$ and $\mu$ in~\eqref{eq:dyncomponents} and the output readings $y_R$, $y_p$ in~\eqref{eq:outcomponents} are
corrupted by sensing noise. For the former we employ the sensor model typically used in aided navigation design~\cite{Farrell}
\begin{equation}
 \begin{aligned}
  \tilde{\omega}&=\omega+\nu_\omega \\
  \tilde{\mu}&=\mu+\nu_\mu
 \end{aligned}
\label{eq:inputsensornoisemodels}
 \end{equation}
where $\nu \sim \mathcal{N}(0,\sigma^2)$ terms represent additive Gaussian white noise vectors whose covariance can be directly identified from logged sensor
data. Remark the noise vectors $\nu_\omega$ and $\nu_\mu$ are expressed in coordinates of the body-fixed frame. The noise models for
scan matching outputs~\eqref{eq:outcomponents} are not standard and will be derived in Section~\ref{sec:ICPaiding}.

\subsection{Geometry of $SO(3)$}
\label{sec:SO3geom}

The special orthogonal group $SO(3)$ is a Lie group. For any Lie group $G$, there exists an exponential map $\exp:\mathfrak{g} \to G$ which maps elements
of the Lie algebra $\mathfrak{g}$ to the Lie group $G$. It is known (e.g.~\cite[p.~519]{Lee}) that for any $X \in
\mathfrak{g}$, $(\exp\, X)^{-1}=\exp(-X)$, that $\exp$ is a smooth map from $\mathfrak{g}$ to $G$, and that $\exp$ restricts to a diffeomorphism from some
neighborhood of $0$ in $\mathfrak{g}$ to a neighborhood of the
identity element $e$ in $G$. The last fact means there exists in a neighborhood $U$ of $e$ an inverse smooth map $\log:G\to \mathfrak{g}$ such that
$\exp \circ \log(g)=g$, $\forall g\in U$.

For the particular case of $SO(3)$, the associated Lie algebra $so(3)$ is the set of $3\times 3$ skew-symmetric
matrices $\xi:=S(x), x\in \mathbb{R}^3$ and $\exp:so(3)\to SO(3)$ is the matrix exponential
\begin{equation*}
 \exp \xi =I+\xi+\frac{1}{2!} \xi^2+\cdots
\end{equation*}
while $\log:SO(3) \to so(3)$ is given by
\begin{equation*}
 \log R = \alpha S(\beta) \qquad \text{where} \qquad \alpha: 2\cos \alpha +1 = \text{trace}(R) \quad \text{and} \quad
S(\beta)=\frac{1}{2\sin\alpha}
(R-R^T)
\end{equation*}
Now consider the special case of $R\in SO(3)$ close to $I$. Since the exponential map restricts to a diffeomorphism from a neighborhood of zero in $so(3)$
to a neighborhood of identity in $SO(3)$, $\xi\in so(3)$ is close to the matrix $0_{3\times
3}$ such that $\xi^2$ and higher-order terms in $R \equiv \exp \xi$ are negligible and thus
\begin{equation}
R \approx I + \xi, \qquad R\in SO(3)\text{ close to }I
\label{eq:Rapprox}
\end{equation}
Since $(\exp \xi)^{-1}=\exp(-\xi)$ we also have
\begin{equation}
R^{-1} \approx I - \xi, \qquad R\in SO(3)\text{ close to }I
\label{eq:Rapproxminus}
\end{equation}
Still for $R$ close to $I$, $\text{trace}(R)\approx 3$ such that $\alpha \approx 0$ and so $\log R \approx (R-R^T)/2$. We define the
projection map
\begin{equation*}
\pi:SO(3) \to so(3),\, \pi(R)=\frac{R-R^T}{2}
\end{equation*}
which is defined everywhere but is the inverse of $\exp$ only for $R\equiv\exp \xi$ close to $I$. In this case $\pi(R)\approx \xi$
and~\eqref{eq:Rapprox} becomes
\begin{equation}
R - I \approx \pi(R), \qquad R\in SO(3)\text{ close to }I
\label{eq:RminusIapprox}
\end{equation}
Approximations~\eqref{eq:Rapprox},~\eqref{eq:Rapproxminus} and~\eqref{eq:RminusIapprox} will be employed in the sequel.

\section{Depth Camera Aiding}
\label{sec:visionaiding}

\subsection{Camera Hardware}
\label{sec:kincam}

Our test robot is equipped with a Kinect depth camera, a low-cost (currently around $100$ \euro)
gaming peripheral sold by Microsoft for the XBox 360
console. The Kinect's sensing technology is
described in~\cite{KM}. The unit 
employs an infrared laser to project a speckle pattern ahead of itself,
whose image is read back using an infrared camera offset from the laser projector. By correlating the acquired image with a stored reference image
corresponding to a known distance, the Kinect computes a disparity map (standard terminology in stereo camera vision) of the scene which is employed to
construct a 3D point
cloud of the environment in the robot-fixed frame. The disparity maps computed by the Kinect, corresponding to
individual pixels of the IR camera image, are available as $640\times 480$ images at a rate of $30\text{ Hz}$. Physically the IR camera has a total angular
field of view of
$57^\circ$
horizontally and
$43^\circ$ vertically, such that the constructed point cloud is fairly ``narrow'' as compared to time-of-flight scanning laser units such as the Hokuyo
UTM-30LX ($270^\circ$) or the Velodyne HDL-32E ($360^\circ$). The maximum depth ranging limit of the Kinect is $5\text{ m}$, in
contrast to the UTM-30LX ($30\text{ m}$) and HDL-32E ($70\text{ m}$). The Kinect can only be utilized under indoor lighting conditions. But conversely, 
the Kinect is much less expensive than the Hokuyo (about $4500$ \euro)
and Velodyne (about $22000$ \euro) units and unlike the scanning units is not prone to point cloud deformation at high vehicle speeds. The scan matching 
algorithms developed for Kinect point clouds are adaptable to these units.

A comprehensive analysis of the Kinect's accuracy and precision is carried out in~\cite{KO12}, showing that point cloud measurements are affected by
both noise
and resolution errors which grow significantly with distance from the camera. Consequently the recommended usable range interval is
$1\leq z \leq 3\text{ m}$.

\subsection{ICP Algorithm}
\label{sec:ICP}

The ICP algorithm~\cite{BM92,CM92} is an iterative procedure for finding the optimum rigid-body transformation $(\delta R,\delta T)\in SO(3)
\times \mathbb{R}^3$ between two sets of points in $\mathbb{R}^3$ (clouds) $\{ a_i \}$ and $\{ b_i \}$, which do not necessarily have equal number of
entries. Following the taxonomy in~\cite{RL01} an ICP algorithm consists of the following (iterated) sequence of steps:
\begin{enumerate}
 \item \emph{Select} source points from one or both clouds. We select a total of $3000$ points (about $1\%$ of the available) from both
clouds and located away from edges, compute their corresponding unit normals $\{ n_i \}$ by fitting a plane through neighboring
points~\cite{KAWB09}, and
employ the strategy of normal-space sampling~\cite{RL01}.
 \item \emph{Match} the source points with those in the other cloud(s) and reject poor matches. We employ a nearest-neighbor search accelerated by a $k-d$ 
tree~\cite{FBF77},
then reject pairs whose point-to-point distance exceeds the threshold value of $0.25\text{ m}$ or whose associated normals form an angle larger than
$45^\circ$.
 \item \emph{Minimize} the error cost function. We choose the point-to-plane ICP variant~\cite{CM92}
\begin{equation}
f(\delta R, \delta T)=\sum_{i=1}^N \big[ ( \delta R a_i + \delta T - b_i )\cdot n_i \big]^2
\label{eq:costICPpn}
\end{equation}
and apply linearization to solve~\eqref{eq:costICPpn} as explained below.
\item \emph{Terminate} if convergence criteria met, else goto 1). For simplicity we do not employ an
early stop condition and always run $25$ iterations of the ICP algorithm.
\end{enumerate}
The heart of the ICP algorithm lies in the way the matching step is done, as the points of the two clouds are arbitrarily labeled and do not initially match 
correctly. Linearization is justified provided $\delta R$ is close to $I$, equivalent to clouds
$\{a_i\}$ and $\{b_i\}$ starting off close to each other. As explained in Section~\ref{sec:ICPaiding} the closeness assumption is valid due to
pre-aligning of the two clouds using an estimate of pose obtained at the prediction step of the overall filter. We thus
employ linearization~\eqref{eq:Rapprox} in~\eqref{eq:costICPpn} to obtain
\begin{equation}
f(x) = \sum_i \big[ (  x_R \times a_i + x_T + a_i - b_i )\cdot n_i \big]^2 = \sum_i \big[   (a_i \times n_i)\cdot x_R + n_i
\cdot x_T + n_i \cdot (a_i - b_i)  \big]^2
\label{eq:costICPpnlin}
\end{equation}
where we have used the scalar triple product circular property $(a \times b)\cdot c = (b \times c)\cdot a$. In the linearized context, minimizing the
ICP cost function~\eqref{eq:costICPpn} is equivalent to minimizing~\eqref{eq:costICPpnlin} $f:\mathbb{R}^6 \to \mathbb{R}$ by choice of $x=[x_R
\quad x_T]^T$. We define the terms
\begin{equation*}
 H_i:=\begin{bmatrix} (a_i \times n_i)^T & n_i^T \end{bmatrix} \qquad y_i:=n_i^T (a_i - b_i)
\end{equation*}
to rewrite~\eqref{eq:costICPpnlin} as
\begin{equation}
f(x) = \sum_i \big[ H_i x + y_i   \big]^2
\label{eq:costICPpnlinalt}
\end{equation}
The minimum of~\eqref{eq:costICPpnlinalt} $f(x)$ is achieved at $x$ for which $\partial f/\partial x=0$ since $\partial^2 f/\partial x^2=2A \geq 0$ everywhere 
with 
\begin{equation*}
\begin{aligned}
 A&:= \sum_i (H_i)^T H_i = \sum_i \begin{bmatrix}  (a_i \times n_i) (a_i \times n_i)^T & (a_i \times n_i) n_i^T \\ n_i (a_i \times n_i)^T & n_i n_i^T
\end{bmatrix}\\
b&:=\sum_i (H_i)^T y_i = \sum_i \begin{bmatrix}  (a_i \times n_i) n_i^T (a_i - b_i) \\ n_i  n_i ^T (a_i -
b_i) \end{bmatrix}
\end{aligned}
\end{equation*}
This minimum is found by solving the linear system of equations $ A x = -b$. 
Then up to second order terms, $(R,p)=(I+S(x_R),x_p)$ is the rigid-body transformation which minimizes
the point-to-plane error~\eqref{eq:costICPpn}.

\subsection{Covariance of ICP output}
\label{sec:ICPcov}

As stated in Section~\ref{sec:kincam} the measured point clouds $\{a_i\}$ and $\{b_i\}$ are affected by sensor noise and resolution errors. We
require a method to estimate the covariance (uncertainty) of the rigid-body
transformation obtained by running the ICP algorithm between measured point clouds. 

We continue to assume $\{a_i\}$ and $\{b_i\}$
start
close to each other (the open-loop integration of the motion sensors allows to pre-align the clouds). We can thus model the (point-to-plane) ICP as a linear least-squares estimator minimizing~\eqref{eq:costICPpnlinalt}
resp.~\eqref{eq:costICPpnlin}. Let $x_{ICP}$ denote the estimate computed by the ICP from noisy point cloud data $\{a_i\}$ and $\{b_i\}$ and $x^\ast$ the true
transformation. The associated covariance is
\begin{equation}
 \text{cov}(x_{ICP})= E\left\langle \left(x_{ICP}-x^\ast\right) \left(x_{ICP}-x^\ast\right)^T \right\rangle
\label{eq:covICP}
 \end{equation}
Based on the linear least-squares cost function~\eqref{eq:costICPpnlinalt} we define the residuals as the error purely due to sensor noise, that is, the error
between each measured point and the original point transformed through the \emph{true} transformation:
\begin{equation}
r_i := H_i x^\ast + y_i
\label{eq:wiresidual}
\end{equation}
The estimate $x_{ICP}$
minimizing~\eqref{eq:costICPpnlinalt} was already shown to be $
x_{ICP}=-A^{-1}b=-\left[ \sum_i (H_i)^T H_i \right]^{-1} \sum_i (H_i)^T y_i$. 
Rewrite this using~\eqref{eq:wiresidual} as $
x_{ICP}=\left[ \sum_i (H_i)^T H_i \right]^{-1} \sum_i (H_i)^T (H_i x^\ast - r_i ) = x^\ast - A^{-1} \sum_i (H_i)^T  r_i$ 
and~\eqref{eq:covICP} becomes
\begin{equation}
\begin{aligned}
\text{cov}(x_{ICP}) &= E\Bigg\langle \bigg( -A^{-1} \sum_i (H_i)^T  r_i \bigg) \bigg( -A^{-1}\sum_i (H_i)^T  r_i \bigg)^T \Bigg\rangle \\
&= \bigg[ \sum_i (H_i)^T H_i \bigg]^{-1} \sum_i \sum_j \Big(
(H_i)^T E \langle r_i r_j \rangle
H_j \Big) \bigg[ \sum_i (H_i)^T H_i \bigg]^{-1}
\end{aligned}
\label{eq:covICPfull}
\end{equation}
In order to evaluate~\eqref{eq:covICPfull} we need to assume a noise model on the residuals $r_i$ in~\eqref{eq:wiresidual}. Expanding the right-hand side using
the definitions of $H_i$ and $y_i$ we have
\begin{equation*}
 \begin{aligned}
  r_i &= (a_i \times n_i)^T x_R^\ast + n_i^T x_T^\ast + n_i^T(a_i-b_i) 
  &=  \left[ (x_R^\ast  \times a_i ) + x_T^\ast  + (a_i-b_i) \right] \cdot n_i := w_i \cdot n_i
 \end{aligned}
\end{equation*}
where $w_i \in \mathbb{R}^3$ represents the post-alignment error of the $i^\text{th}$ point pair due (only) to the presence of sensor noise in point
clouds $\{a_i\}$ and $\{b_i\}$. The residual $r_i=w_i \cdot n_i = n_i^T w_i$ now represents the projection of $w_i$ and~\eqref{eq:covICPfull} becomes 
\begin{equation}
\text{cov}(x_{ICP})= \bigg[ \sum_i (H_i)^T H_i \bigg]^{-1} \sum_i \sum_j \Big(
(H_i)^T n_i^T E \langle w_i w_j^T \rangle n_j
H_j \Big) \bigg[ \sum_i (H_i)^T H_i \bigg]^{-1}
\label{eq:covICPfull2}
\end{equation}
The classical Hessian method~\cite{BB03,HBG12,BBDG14} consists of assuming the post-alignment errors $w_i$ are independent and identically
normally isotropically distributed with standard deviation $\sigma$. 
Under these assumptions $E \langle w_i w_j^T \rangle = E \langle w_i \rangle E \langle w_j^T \rangle =0$, $i\neq j$ and the double sum
in~\eqref{eq:covICPfull2} reduces to a single sum ($n_i^T n_i=1$ for unit normals):
\begin{equation*}
\text{cov}(x_{ICP}) = \bigg[ \sum_i (H_i)^T H_i \bigg]^{-1} \sigma^2  \sum_i
(H_i)^T H_i \bigg[ \sum_i (H_i)^T H_i \bigg]^{-1} = \sigma^2 \bigg[ \sum_i (H_i)^T H_i \bigg]^{-1}
\end{equation*}
This is precisely the covariance of a linear unbiased estimator using observations with additive white Gaussian noise c.f.~\cite[p.~85]{Kay}.
However, there is a catch: for a Kinect sensor with $N\approx 300\,000$ points per cloud and $\sigma \approx 1\text{ cm}$ for depth range $1\leq z \leq 3\text{
m}$~\cite[Fig.~10]{KO12}, the above expression yields a covariance matrix with entries on the order of nanometers, a completely overoptimistic result for
the low-cost Kinect sensor. This result stems from the independence assumption which makes the estimator converge as $1/\sqrt{N}$ where $N$ is the (high) 
number of points. 

We thus need to consider a more complete error model for the residuals. In addition to random noise on the residuals whose contribution to
$\text{cov}(x_{ICP})$ rapidly becomes negligible as the number of point pairs $N$ increases, the Kinect exhibits resolution errors on the
order of $\delta \approx 1\text{ cm}$ for $1 \leq z \leq 3\text{ m}$~\cite[Fig.~10]{KO12} due to quantization errors incurred during IR image correlation to a
reference image as discussed in Section~\ref{sec:kincam}. These resolution errors are non-zero mean and are not independent of each other and actually 
constitute the dominant limitation in accuracy. As shown in our preliminary work~\cite{BBG14}, this leads to the following approximation of the covariance 
matrix~\eqref{eq:covICPfull2}:
\begin{equation}
 \text{cov}(x_{ICP}) \approx \delta^2 \frac{N}{N_p} \bigg[ \sum_i (H_i)^T H_i \bigg]^{-1} =  \delta^2 \frac{N}{N_p} \left(  \sum_{i=1}^N \begin{bmatrix}  (a_i \times n_i) (a_i
\times
n_i)^T & (a_i \times n_i) n_i^T
\\
n_i (a_i \times n_i)^T & n_i n_i^T
\end{bmatrix} \right)^{-1}
\label{eq:pncovfinal}
\end{equation}
where $N_p=3$ is the number of plane ``buckets'' used for normal-space sampling~\cite{RL01} in Step 1 of the ICP algorithm, c.f.~Section~\ref{sec:ICP}. Note 
that the covariance matrix correctly reflects the observability of the environment: for instance if the environment consists of a single plane, all $n_i$'s are 
identical leading to rank deficiency of the matrix to be inverted, thus the covariance is infinite along directions parallel to the plane. Also note that the 
overall ICP variance is on the order of the Kinect's precision $\delta^2$ in the case of full observability, as expected.

\subsection{Scan Matching as Measured Output}
\label{sec:ICPaiding}

Suppose a 3D map of the environment made up of point clouds is already available, built either by the robot or from lasergrammetry. Assuming the scan from 
the on-board depth camera and the map contain a set of identical features, the robot's pose with respect to the map can be identified through an ICP 
algorithm. However, due to sensor noise and quantization effects, the computed pose is noisy. 

Denote the initial estimate of the robot's pose (obtained from numerical integration of the vehicle dynamics~\eqref{eq:dyncomponents}) as ($R^-,p^-)$. 
This estimate is used to pre-align the two scans (the current scan and the map) in
the body-fixed frame; this is
required since the ICP does not guarantee global convergence, and in fact can easily get stuck at a local minimum. The robot's true pose can then be expressed 
up to second order terms as ($R^-(I+S(x_R^*)),R^-x_p^*+p^-)$, that is the vector $(x_R^*,x_p^*)\in\mathbb{R}^6$ is defined as the discrepancy of the initial 
estimation ($R^-,p^-)$ and the true pose ($R^*,p^*)$ projected in the Lie algebra $\mathbb{R}^6$ and is the vector to be estimated by the ICP algorithm. 

Based on Section~\ref{sec:ICPcov}, we have seen that the ICP output writes $x_{ICP}=(x_R^*,x_p^*)+(\nu_R,\nu_p)$ where the covariance of the vector 
$(\nu_R,\nu_p)\in \mathbb R^6$ is given by \eqref{eq:covICPfull2} and can be approximated by the simple to compute expression 
\eqref{eq:pncovfinal}. Using the linearity of the map $S$, we see that $I+S((x_{ICP})_R+\nu_R)=I+S(x_R^*)+S(\nu_R)$ and so the pose output from 
scan matching is (up to second-order terms)
\begin{equation}
\begin{pmatrix}
 \tilde{y}_R \\
 \tilde{y}_p
\end{pmatrix}
=
\begin{pmatrix}
 R^* \\
 p^*
\end{pmatrix}
+ \begin{pmatrix}
R^* S(\nu_R) \\
R^* \nu_p
\end{pmatrix}
\label{eq:finalYmICPb}
\end{equation}
The above is a noisy version of~\eqref{eq:outcomponents}, where the noise covariance $\text{cov}(\upsilon)$, $\upsilon:=[\nu_R\quad \nu_p]^T$ is (approximately 
but efficiently) computed by~\eqref{eq:pncovfinal}.

\section{State Estimator Design}
\label{sec:estimatordesign}

\subsection{Invariant Observer}
\label{sec:invobsdesign}

We first design an invariant observer for the nominal (noise-free) system~\eqref{eq:dyncomponents},~\eqref{eq:outcomponents} by following the
constructive method in~\cite{BMR08,BMR09}; a
tutorial presentation is available in~\cite{BL13} and so the calculations will be omitted. We consider the case where the Special Euclidean Lie group 
$SE(3) = SO(3)\times\mathbb{R}^3$ acts on the $SE(3)$ state of~\eqref{eq:dyncomponents} by left translation, which physically represents applying
a constant rigid-body transformation $(R_0,p_0)$ to ground-fixed frame vector coordinates. This leads to the nonlinear invariant observer
\begin{equation}
\begin{pmatrix}
  \dot{\hat{R}} \\ \dot{\hat{p}}
 \end{pmatrix}= \begin{pmatrix}
   \hat{R} S(\omega) \\ \hat{R} \mu
  \end{pmatrix}
+ \hat{R} \begin{pmatrix} S \big(  L_R^R E_R + L_p^R E_p \big) \\ L_R^p E_R + L_p^p E_p \end{pmatrix}
\label{eq:invleftobs}
\end{equation}
where $L_R^R$, $L_p^R$, $L_R^p$, $L_p^p$ are $\mathbb{R}^{3\times 3}$ gain matrices and $$E_R:=-S^{-1}(\pi(\hat{R}^T y_R)),\quad E_p=\hat{R}^T (\hat{p} - y_p)$$ 
are the invariant output error column vectors. Standard computations in the framework of symmetry-preserving observers show the associated invariant estimation 
errors $\eta_R=R^T \hat{R}$, $\eta_p=R^T(\hat{p}-p)$ have dynamics 
\begin{equation}
 \begin{aligned}
\dot{\eta}_R &= -S(\omega)  \eta_R + \eta_R S(\omega) + \eta_R S\big(  L_R^R E_R + L_p^R E_p \big)\\
\dot{\eta}_p &= -S(\omega) \eta_p + \eta_R \mu - \mu + \eta_R \big( L_R^p E_R + L_p^p E_p \big)
 \end{aligned}
 \label{eq:ieed}
\end{equation}
and stabilizing the (nonlinear) dynamics~\eqref{eq:ieed} to $\overline{\eta}=I$ by choice of gains $L$ leads to an asymptotically stable nonlinear
observer~\eqref{eq:invleftobs}. The stabilization process is simplified by~\eqref{eq:ieed} not being dependent on the estimated system state $\hat{x}$; indeed
the
fundamental
feature of the invariant observer is that it guarantees $\dot{\eta} =
\Upsilon(\eta,I(\hat{x},u))$~\cite[Thm.~2]{BMR08} where $I(\hat{x},u)=[\omega \quad \mu]$ is the set of invariants in the present example. We will compute 
the stabilizing gains using the Invariant EKF method in order to handle variable scan matching observability of the environment. 

\subsection{Invariant EKF}
\label{sec:LIEKF}

The Invariant EKF~\cite{Bon07} is a systematic approach to computing the gains $K$ of an invariant observer by linearizing
its
invariant estimation error dynamics. A noisy version of~\eqref{eq:dyncomponents} is readily obtained by accounting for the sensors' noise by expressing 
the measured angular and linear velocities as the noisy vectors $\omega=\tilde\omega+\nu_\omega$ and $\mu=\tilde\mu+\nu_\mu$ and letting
 $Q_\nu:=\text{cov}[\nu_\omega \quad \nu_\mu]^T$ denote the process white noise covariance matrix. The entries of $Q_\nu$ can be
identified directly from logged sensor data,
while $R_\nu := \text{cov}[\nu_R\quad \nu_p]^T $ is computed by \eqref{eq:covICPfull2} (or more simply~\eqref{eq:pncovfinal}). 

Following the IEKF method, we first write a noisy version of the error dynamics \eqref{eq:ieed} where we let $\omega=\tilde\omega-\nu_\omega$ and 
$\mu=\tilde\mu-\nu_\mu$, where $\tilde\omega,\tilde\mu$ denote the noisy inputs read from onboard sensors, and the output is given by \eqref{eq:finalYmICPb}.  
This equation is linearized using the standard methodology of symmetry-preserving observers, i.e. using vectors of $\mathbb R^3$ to denote the orientation error 
by letting $\eta_R:=I+S(\zeta_R)$, $\zeta_R\in\mathbb{R}^3$ by~\eqref{eq:Rapprox} and letting $\eta_p:=\zeta_p \in \mathbb{R}^3$. Up to second order terms in 
$\zeta,\nu$, the noisy version of \eqref{eq:ieed} is approximated by the following linear equation:
\begin{equation*}
 \frac{d}{dt} \begin{pmatrix} \zeta_R \\ \zeta_p \end{pmatrix}
= \begin{pmatrix}
   -S(\tilde{\omega}) & 0 \\ -S(\tilde{\mu}) & -S(\tilde{\omega})
  \end{pmatrix}
\begin{pmatrix} \zeta_R \\ \zeta_p \end{pmatrix}
+ \begin{pmatrix} \nu_\omega \\ \nu_\mu \end{pmatrix}
+\begin{pmatrix}
   L_R^R & L_p^R \\ L_R^p & L_p^p
  \end{pmatrix}
\begin{pmatrix} \zeta_R \\ \zeta_p \end{pmatrix}
- \begin{pmatrix}
   L_R^R & L_p^R \\ L_R^p & L_p^p
  \end{pmatrix}
\begin{pmatrix}
\nu_R \\
\nu_p
\end{pmatrix}
\end{equation*}
The rationale of the IEKF is to tune the gains through Kalman theory in order to minimize at each step the increase in the covariance of the linearized error 
$\zeta$. This is done through the standard Kalman filter equations, letting 
\begin{equation}
 A= \begin{bmatrix}
   -S(\tilde{\omega}) & 0 \\ -S(\tilde{\mu}) & -S(\tilde{\omega})
  \end{bmatrix}, \quad B = \begin{bmatrix}
   -I & 0 \\ 0 & -I
  \end{bmatrix}, \quad C = \begin{bmatrix}
   -I & 0 \\ 0 & -I
  \end{bmatrix}, \quad D = \begin{bmatrix}
   -I & 0 \\ 0 & -I
  \end{bmatrix}, \quad K= \begin{bmatrix}
   L_R^R & L_p^R \\ L_R^p & L_p^p
  \end{bmatrix}
  \label{eq:IEKFmatrices}
\end{equation}
and tuning the gains through the standard Riccati equations in continuous time of the Kalman-Bucy filter
\begin{equation}
\begin{aligned}
\dot{P}&=AP + PA^T - P R_{\nu}^{-1}P + Q_{\nu}\\
K&=-P R_{\nu}^{-1}
\end{aligned}
\label{eq:CTKalmanK}
\end{equation}
where the gains $L$ making up $K$ are employed in the invariant observer~\eqref{eq:invleftobs}.

Remark the IEKF filter matrices $(A,B,C,D)$
are dependent on the system trajectory only through the
$\omega_m$ and $\mu_m$ terms in $A$, and do not depend on the estimates ($\hat R,\hat p$) as in the usual Extended Kalman Filter. The
interest of the Invariant EKF is indeed the reduced dependence of the linearized system on the estimated trajectory of the target system. In our present example 
the $(A,B,C,D)$ matrices are
guaranteed not to
depend on the estimated state, which increases the filter's robustness to poor state estimates and precludes divergence (c.f.~Section~\ref{sec:intro}). 

\subsection{Multiplicative EKF design}
\label{sec:convEKFdes}

For comparison purposes consider a typical~\cite{Farrell} Multiplicative Extended Kalman Filter (MEKF) design for our system. The governing system equations 
are given by~\eqref{eq:dyncomponents} with noise models~\eqref{eq:inputsensornoisemodels} and output model~\eqref{eq:finalYmICPb}:
\begin{equation}
 \begin{aligned}
\dot{R}&= R S(\tilde{\omega}-\nu_\omega) \\
\dot{p}&=R(\tilde{\mu}-\nu_\mu) \\
\begin{bmatrix}
 y_R\\y_p
\end{bmatrix} &= \begin{bmatrix}
 R + RS(\nu_R) \\ p + R\nu_p
\end{bmatrix}
 \end{aligned}
\label{eq:EKFsys}
 \end{equation}
We linearize~\eqref{eq:EKFsys} about a nominal system trajectory $(\hat{R},\hat{p})$. Remark $(R-\hat{R}) \notin SO(3)$ is not a valid linearized system state. 
Instead define the multiplicative attitude error $\Gamma:=\hat{R}^T R \in SO(3)$
such that for $R$ close to $\hat{R}$, $\Gamma$ is close to $I$. By~\eqref{eq:Rapprox} $\Gamma:=I+S(\delta \gamma)$, $\delta \gamma\in\mathbb{R}^3$ and 
so $\hat{R}^T R=I+S(\delta \gamma) \Longrightarrow R - \hat{R} = \hat{R} S(\delta \gamma)$. We define $\delta p = p - \hat{p}$, the output errors $\delta y_R 
:= 
S^{-1}[ \pi(\hat{R}^T y_R) ]$ with $\pi:SO(3) \to so(3)$ from Section~\ref{sec:SO3geom} and $\delta y_p := y_p - \hat{p}$ and obtain the linearized system
\begin{equation}
 \begin{aligned}
  \begin{bmatrix}
   \delta \dot{\gamma} \\ \delta \dot{p}
  \end{bmatrix}
&= \begin{bmatrix}
    -S(\tilde{\omega}) & 0 \\ -\hat{R}S(\tilde{\mu}) & 0
   \end{bmatrix}\begin{bmatrix}
   \delta \gamma \\ \delta p
  \end{bmatrix}
+ \begin{bmatrix}
   -I & 0 \\ 0 & -\hat{R}
  \end{bmatrix} \begin{bmatrix}
  \nu_\omega \\ \nu_\mu \end{bmatrix}\\
  \begin{bmatrix}
   \delta y_R \\ \delta y_p
  \end{bmatrix} &= \begin{bmatrix}
  I & 0 \\ 0 & I \end{bmatrix}\begin{bmatrix}
   \delta \gamma \\ \delta p
  \end{bmatrix} + \begin{bmatrix}
   I & 0 \\ 0 & \hat{R}
  \end{bmatrix} \begin{bmatrix}
  \nu_R \\ \nu_p \end{bmatrix}
 \end{aligned}
 \label{eq:convEKFerrsys}
\end{equation}
an LTV system tractable using the classical Kalman Filter. The resulting $[\delta p \quad \delta \gamma]$ estimate is used to 
update the estimated state of the nonlinear system~\eqref{eq:EKFsys} as follows. Note $S(\delta \gamma)\in so(3)$ corresponds to $\Gamma = 
\hat{R}^T R \in SO(3)$ and by Section~\ref{sec:SO3geom} $\exp:so(3) \to SO(3)$ is the matrix exponential. Thus the estimated states are updated as
\begin{equation*}
 \hat{p}^+ = \hat{p} + \delta p \qquad \text{and} \qquad \hat{R}^+ = \hat{R} \exp S(\delta \gamma)
\end{equation*}

The main difference between the MEKF and the IEKF is that the former linearizes the system dynamics~\eqref{eq:EKFsys} about a nominal trajectory, while the 
latter linearizes the invariant estimation error dynamics~\eqref{eq:ieed} about identity. As discussed in Section~\ref{sec:invobsdesign} 
and~\ref{sec:LIEKF}, the latter dynamics do not depend on the estimated state $\hat{x}$ such that the linearized IEKF system is guaranteed to be robust to poor 
estimates of state. Meanwhile the MEKF cannot make this guarantee and indeed the linearized system
matrices~\eqref{eq:convEKFerrsys} depend on the estimated state via $\hat{R}$. The difference in experimental estimation performance of the two designs will be 
demonstrated in Section~\ref{sec:expvalidation}. 

\section{Experimental Validation}
\label{sec:expvalidation}

\subsection{Hardware platform}
\label{sec:platform}

\begin{figure}[!htb]
   \centering
   \begin{tabular}{c c}
    \includegraphics[height=0.4\linewidth]{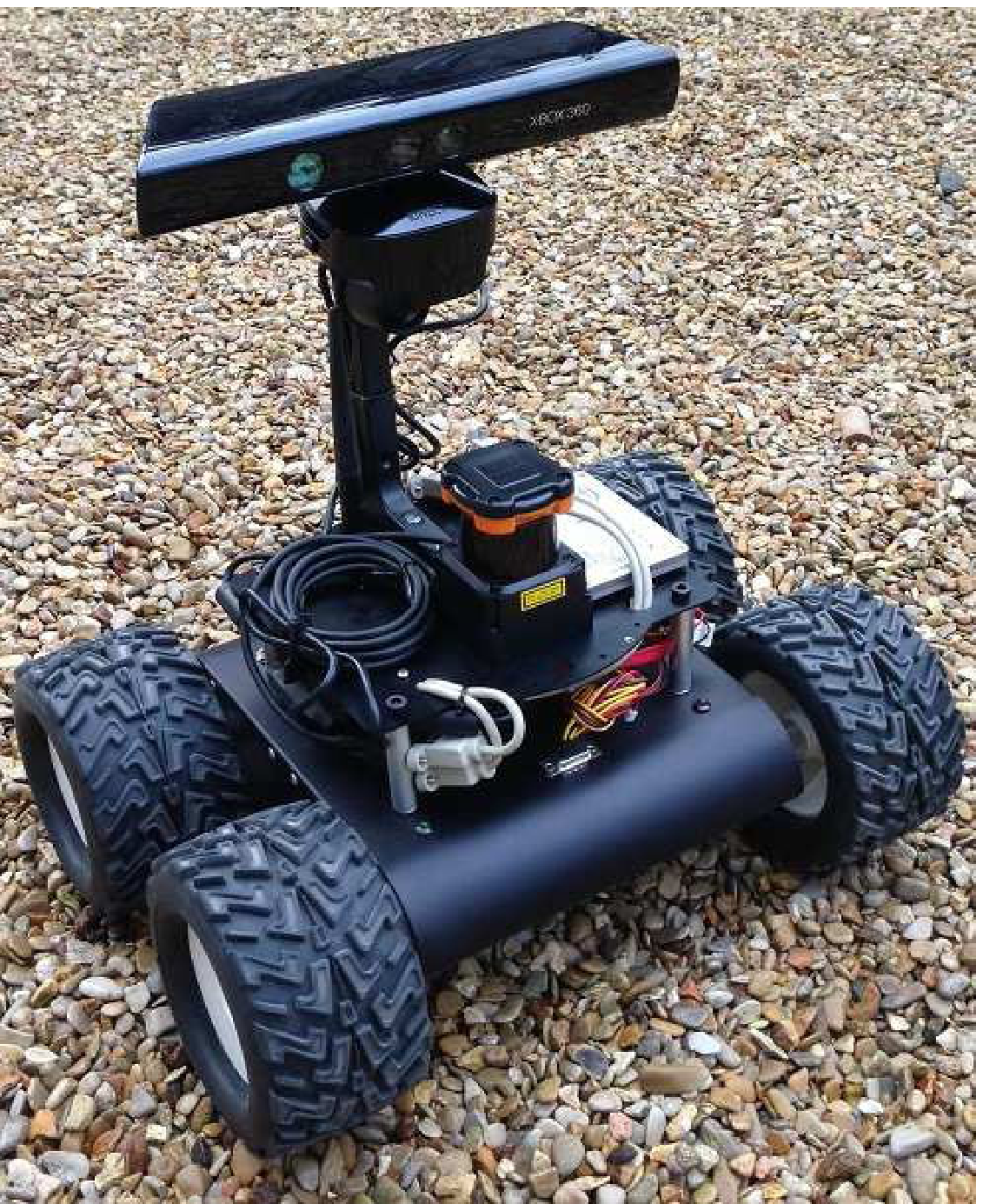} & \includegraphics[height=0.4\linewidth]{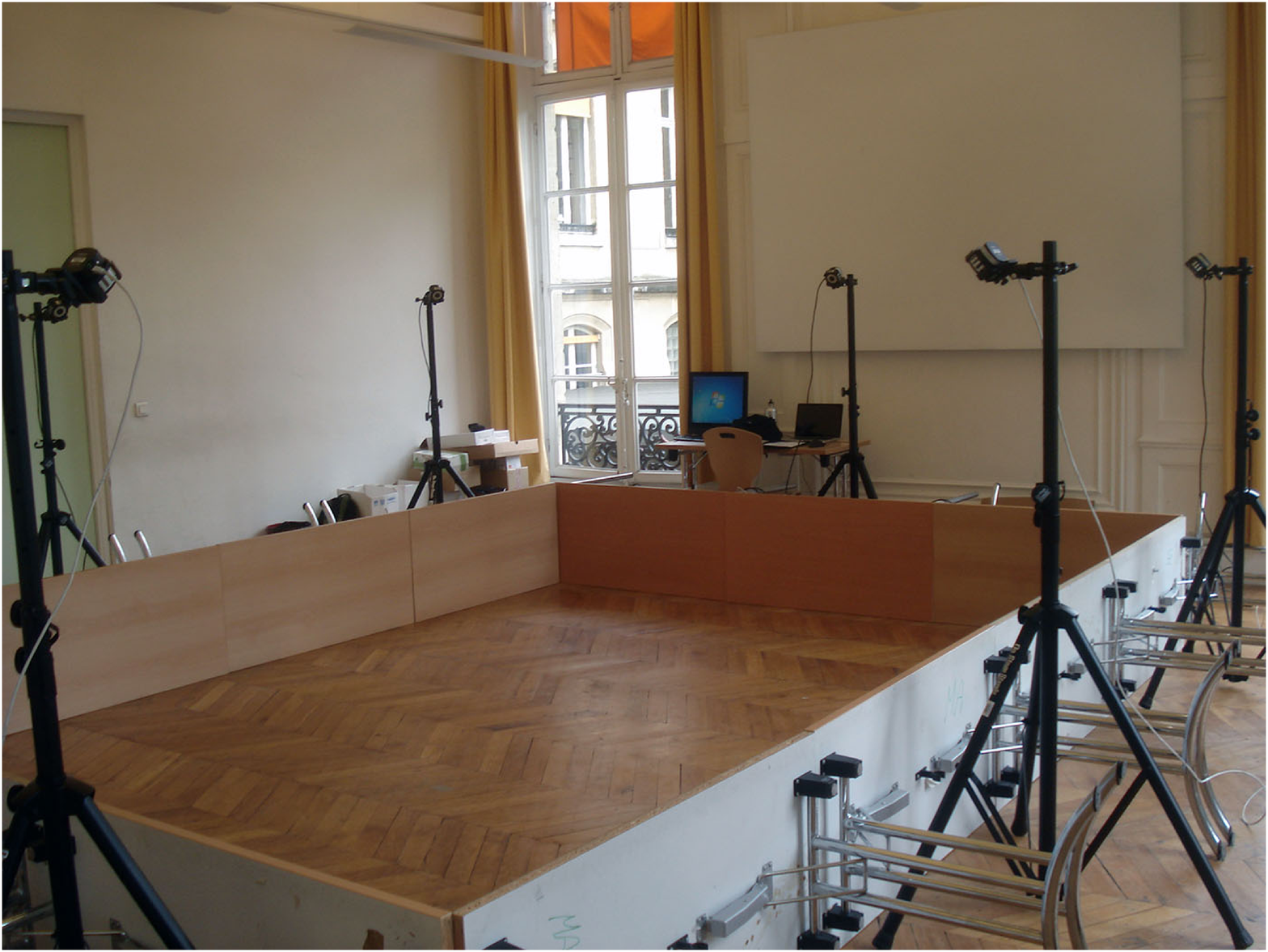}
   \end{tabular}
 \caption{The Wifibot Lab v4 Robot (left); Experiment area with motion-capture system (right)}
\label{fig:wifibotpic}
\end{figure}

The wheeled robot used for our experiments is shown on the left of Figure~\ref{fig:wifibotpic}. The robot is equipped with an Intel Core i5-based
single-board computer running Ubuntu Linux, WLAN 802.11g wireless networking, all-wheel drive via $12\text{ V}$ brushless DC motors, and a Kinect camera 
providing 3-D point
cloud scans of the environment. The experimental testing area is shown on the right side of Figure~\ref{fig:wifibotpic}, and consists of
an open area surrounded by a set of seven S250e cameras employed by an OptiTrack motion capture system to provide a set of ground truth (reference trajectory)
data for
the experiments with sub-millimeter precision at a rate of $120\text{ Hz}$.
The wooden parquet floor seen in Figure~\ref{fig:wifibotpic} is not perfectly flat, causing small oscillations in the vehicle's attitude
and height
which will be visible in the experimental plots. A number of visual landmarks (rectangular boxes) were placed randomly around the
experimental area in order
to provide good scan-matching conditions.

The robot is equipped with independent odometers on the left and right wheels. The two odometer counts are averaged and converted to forward velocity
$\mu_x$ by dividing by an experimentally-identified constant $\kappa_1=788\text{ m}^{-1}$ representing the number of encoder counts per meter of
travel. Assuming zero side
slip as well as zero out-of-plane velocity, we obtain the body-fixed velocity vector $\mu=[\mu_x \quad 0 \quad 0]^T\text{ m/s}$ used in noiseless vehicle
dynamics~\eqref{eq:dyncomponents}. Since the present vehicle is not equipped with a rate gyro, we employ the difference in odometer counts, identified constant
$\kappa_2=0.44\text{ m}$ representing the lateral distance between wheel-ground contact points and $\kappa_1$ to compute in-plane angular velocity
$\omega_z$
then employ the angular velocity vector $\omega=[0 \quad 0 \quad \omega_z]^T\text{ rad/s}$ in~\eqref{eq:dyncomponents}, i.e.~assume zero roll and pitch angular
velocity due to the vehicle being level, with the non-flatness of the floor reflected by additive sensor noise vectors $\nu_\mu$ and $\nu_\omega$. The
covariances of $\nu_\mu$ and $\nu_\omega$ were assigned by first identifying the on-axis variances
$\sigma_{\mu_x}^2=0.01^2\text{ m}^2\text{/s}$ and $\sigma_{\omega_z}^2=0.02^2\text{ rad}^2\text{/s}$ of data logged during respectively constant-velocity
advance and constant-velocity circle
trajectories, then taking $\text{diag}(\text{cov}(\nu_\mu))=[\sigma_{\mu_x}^2 \quad 0.1 \sigma_{\mu_x}^2 \quad 0.1 \sigma_{\mu_x}^2]$ and
$\text{diag}(\text{cov}(\nu_\omega))=[0.1 \sigma_{\omega_x}^2 \quad
0.1 \sigma_{\omega_x}^2 \quad \sigma_{\omega_z}^2]$ on account of the uneven terrain.

Clearly the assumptions in the previous paragraph are both optimistic and ad-hoc, for instance the zero-slip assumption does not fully hold, the noise
covariances are assigned heuristically, and the encoder-derived data is subject to identification errors of $\kappa$ and
quantization effects. This is acceptable for our purposes, however, since we are interested in comparing the experimental performance of two competing filter
designs
more than the absolute accuracy of the estimates. Since we will employ identical sensor data in both designs, we will be able to make a fair comparison
between the two.

For each experiment, the sampling rate of the wheel odometers was set to $50\text{ Hz}$ and the Kinect depth images to $1\text{ Hz}$. The latter is a fairly
slow rate for scan matching and was chosen specifically to test the robustness of each of the two filters. The trajectories were steered in open-loop mode
by setting left and right wheel velocities. The map was built from the initial robot's pose. The resulting sensor data was logged to the on-board memory and 
then run
through both the Invariant EKF and the Multiplicative EKF whose designs were covered throughout
Section~\ref{sec:estimatordesign}. When an image was found to possess a substantial amount of novel information compared to the existing map it was aggregated 
to the existing map. For fairness of comparison the ICP algorithm and associated parameters e.g.~number of point pairs and
rejection criteria (c.f.~Section~\ref{sec:ICP}) were identical among the filters.

\subsection{First Experiment}

\begin{figure}[!hbt]
\begin{center}
\begin{tabular}{c c}
\includegraphics[width=0.4\linewidth]{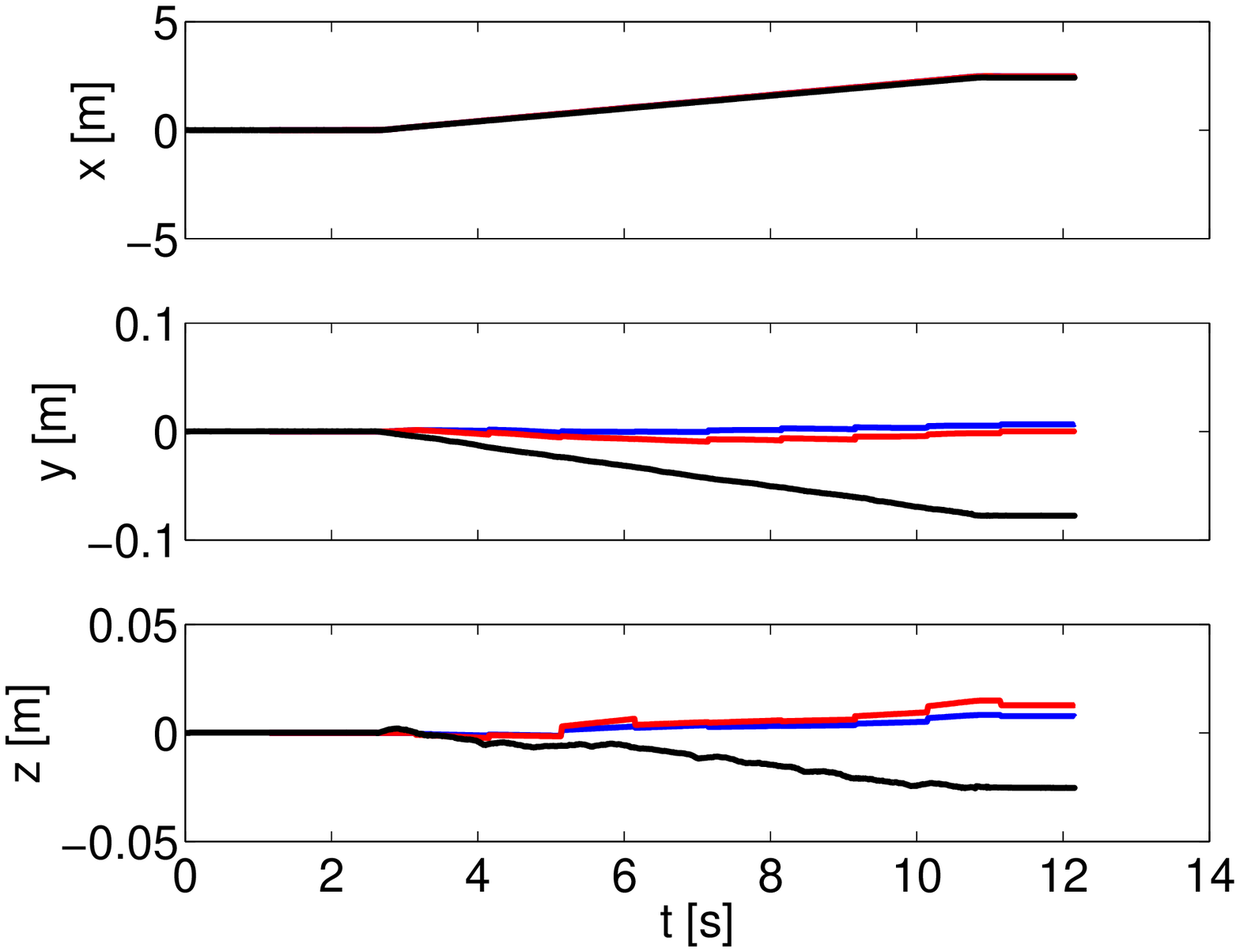} & \includegraphics[width=0.4\linewidth]{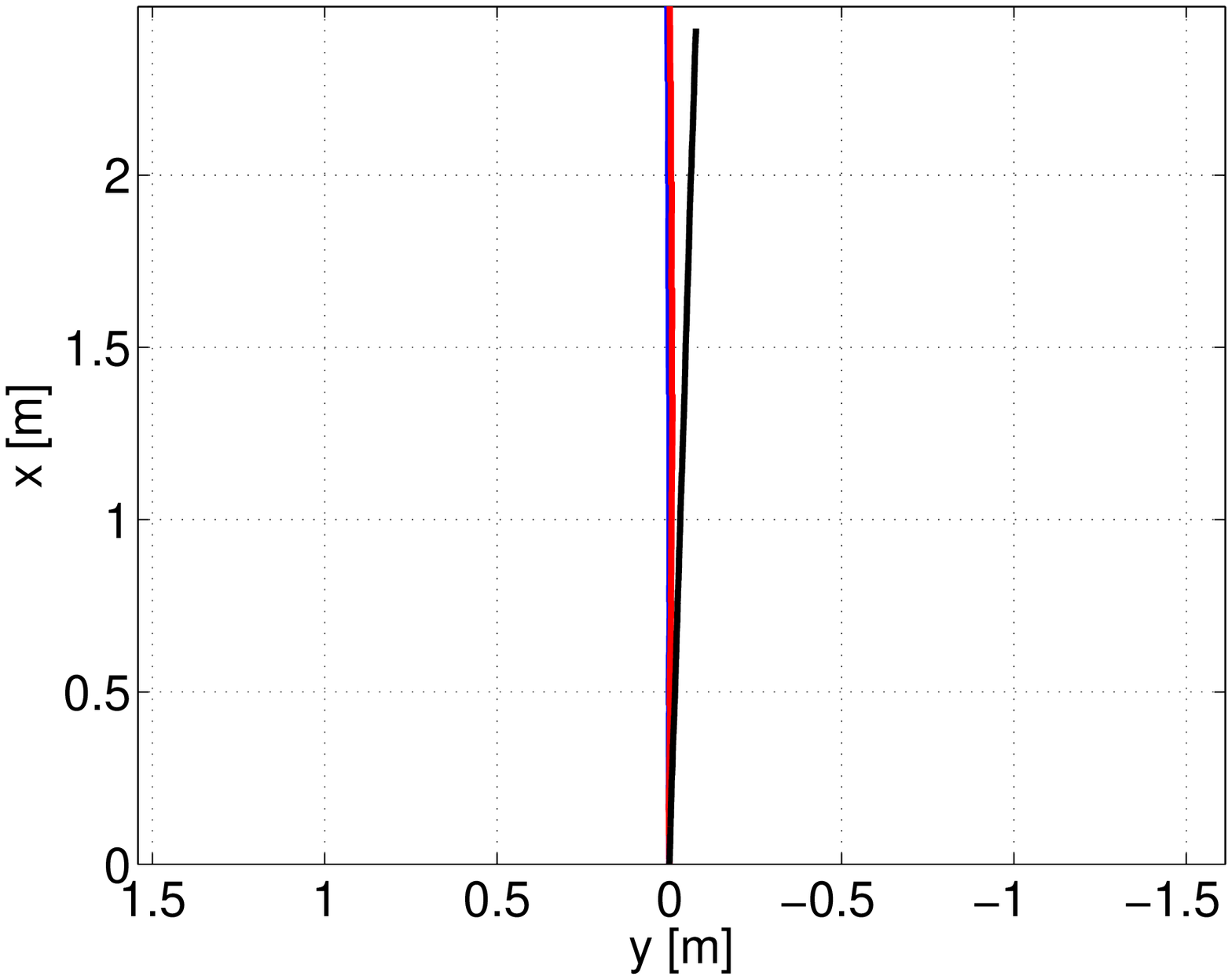} \\
\includegraphics[width=0.4\linewidth]{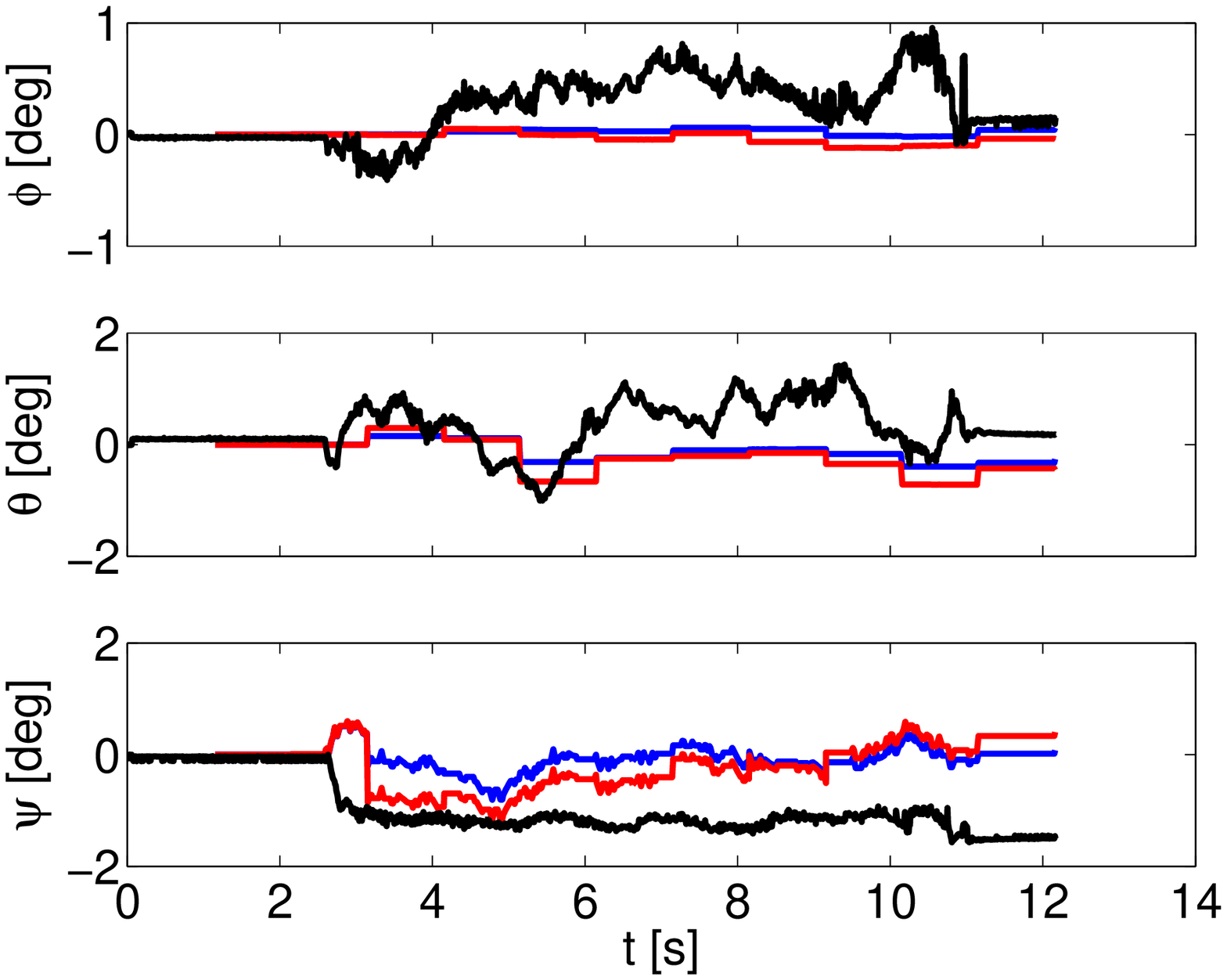} & \includegraphics[width=0.4\linewidth]{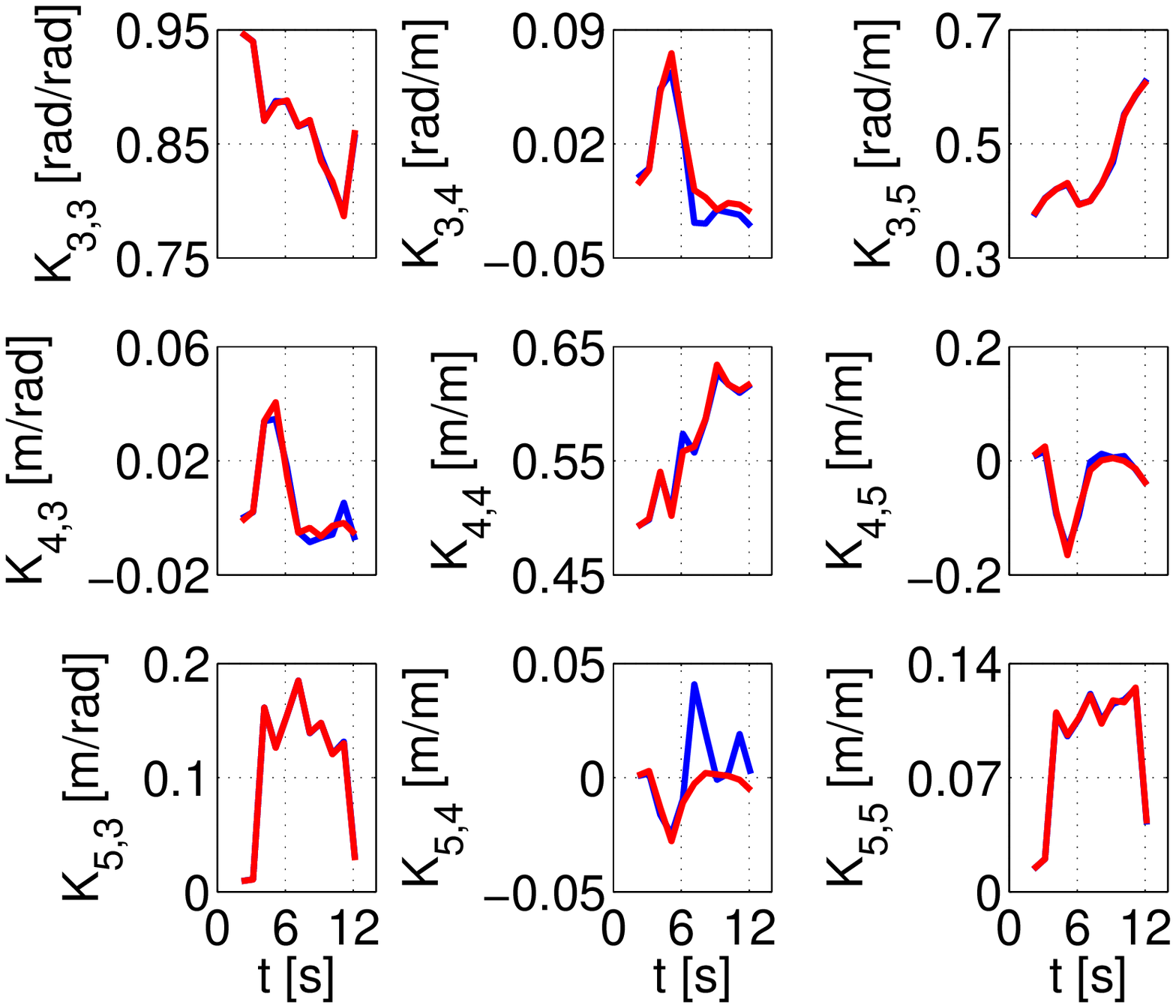}
\end{tabular}
\caption{Linear trajectory experiment: IEKF (\textcolor[rgb]{0,0,1}{\protect\rule[0.25ex]{0.5cm}{2pt}}), MEKF 
(\textcolor[rgb]{1,0,0}{\protect\rule[0.25ex]{0.5cm}{2pt}}), Ground truth (\textcolor[rgb]{0,0,0}{\protect\rule[0.25ex]{0.5cm}{2pt}})}
\label{fig:exp1}
\end{center}
\end{figure}

In the first experiment, the Wifibot starts stationary near an edge of the bounding area wall, then advances in a straight line towards the opposite wall,
where it stops. The position and attitude (converted to Euler angles) estimates from the IEKF and MEKF designs are plotted against
the OptiTrack ground truth in Figure~\ref{fig:exp1}. For visualization purposes Figure~\ref{fig:exp1} also provides an overhead view of
the positions, plus the values of the $3\times 3$ subset of the Kalman gain matrix $K$ acting on planar position and heading angle error states via
their corresponding errors. 

From Figure~\ref{fig:exp1} we see that the IEKF and MEKF designs perform nearly the same. In the final stationary configuration, the
IEKF exhibits an in-plane error of $(\Delta x,\Delta y,\Delta \psi)=(6.2\text{ cm},8.5\text{ cm},1.6^\circ)$ from the ground truth, while for the MEKF this
error is $(6.0\text{ cm},7.8\text{ cm},1.9^\circ)$ --- the difference being within the uncertainty of the system. The RMS of the vector of
errors between estimated trajectories and ground truth throughout the full experiment are $\text{RMS}(\Delta x,\Delta y,\Delta
\psi)=(3.8\text{ cm}, 4.9\text{ cm}, 1.1^\circ)$ for the IEKF and $(3.6\text{ cm}, 4.4\text{ cm}, 1.0^\circ)$ for the MEKF, again within uncertainty. In
addition the Kalman gain matrix entries are seen to behave nearly the same. This almost-identical performance is as expected because for a linear
trajectory both filters reduce to a linear Kalman filter. We will be considering a non-linear robot trajectory in the next experiment.

Note that the OptiTrack data reflects the unevenness of the wooden floor mentioned in Section~\ref{sec:platform}, which causes the robot to sway with
an order-of-magnitude amplitude of $1^\circ$ in the roll and pitch axes and heave with amplitude $1\text{ cm}$ in the vertical axis. This acts as an exogenous
disturbance to the system, and as previously discussed is accounted for by the additive process noise vectors $\nu_\mu$ and $\nu_\omega$.

\subsection{Second Experiment}

\begin{figure}[!hbt]
\begin{center}
\begin{tabular}{c c}
\includegraphics[width=0.4\linewidth]{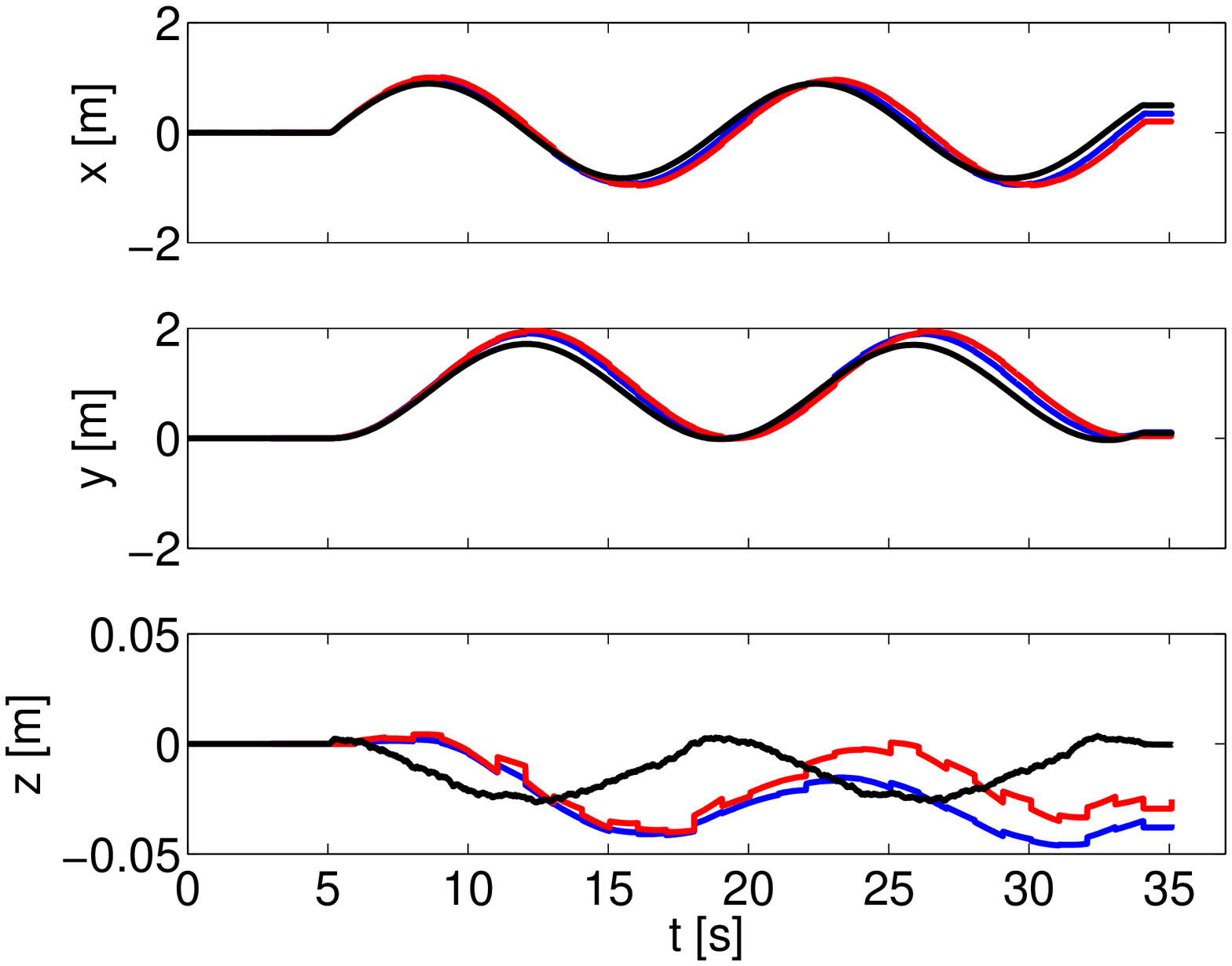} & \includegraphics[width=0.4\linewidth]{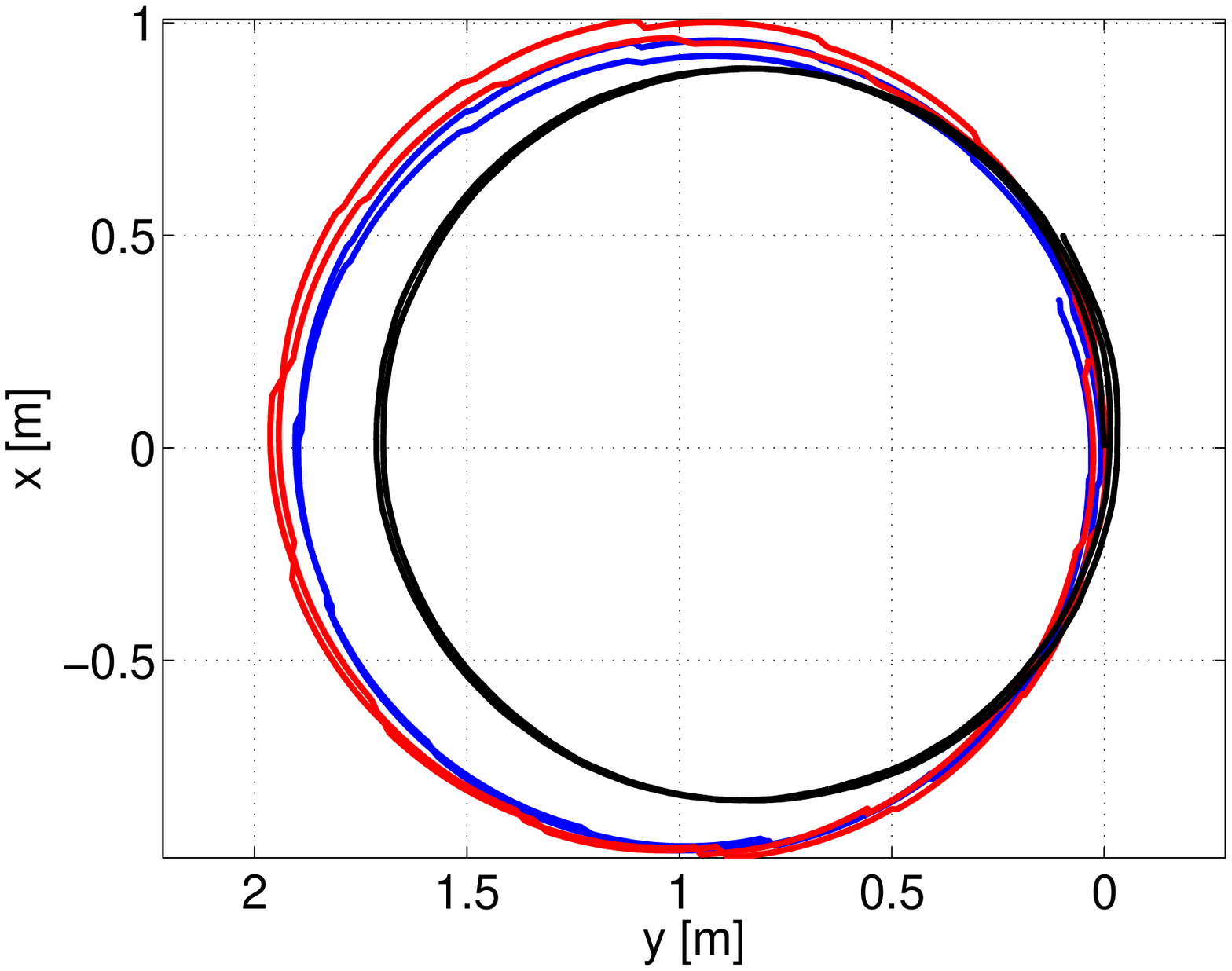} \\
\includegraphics[width=0.4\linewidth]{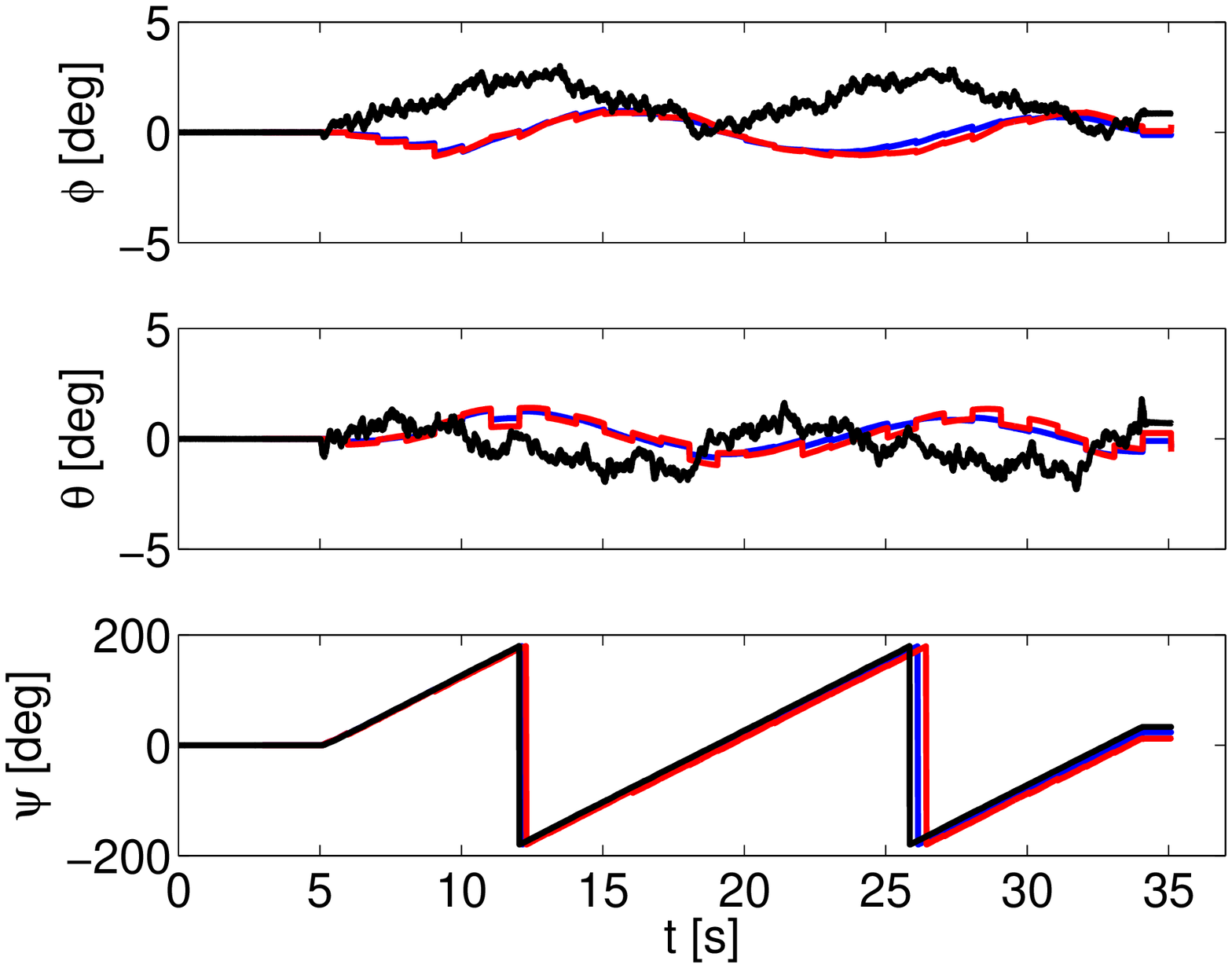} & \includegraphics[width=0.4\linewidth]{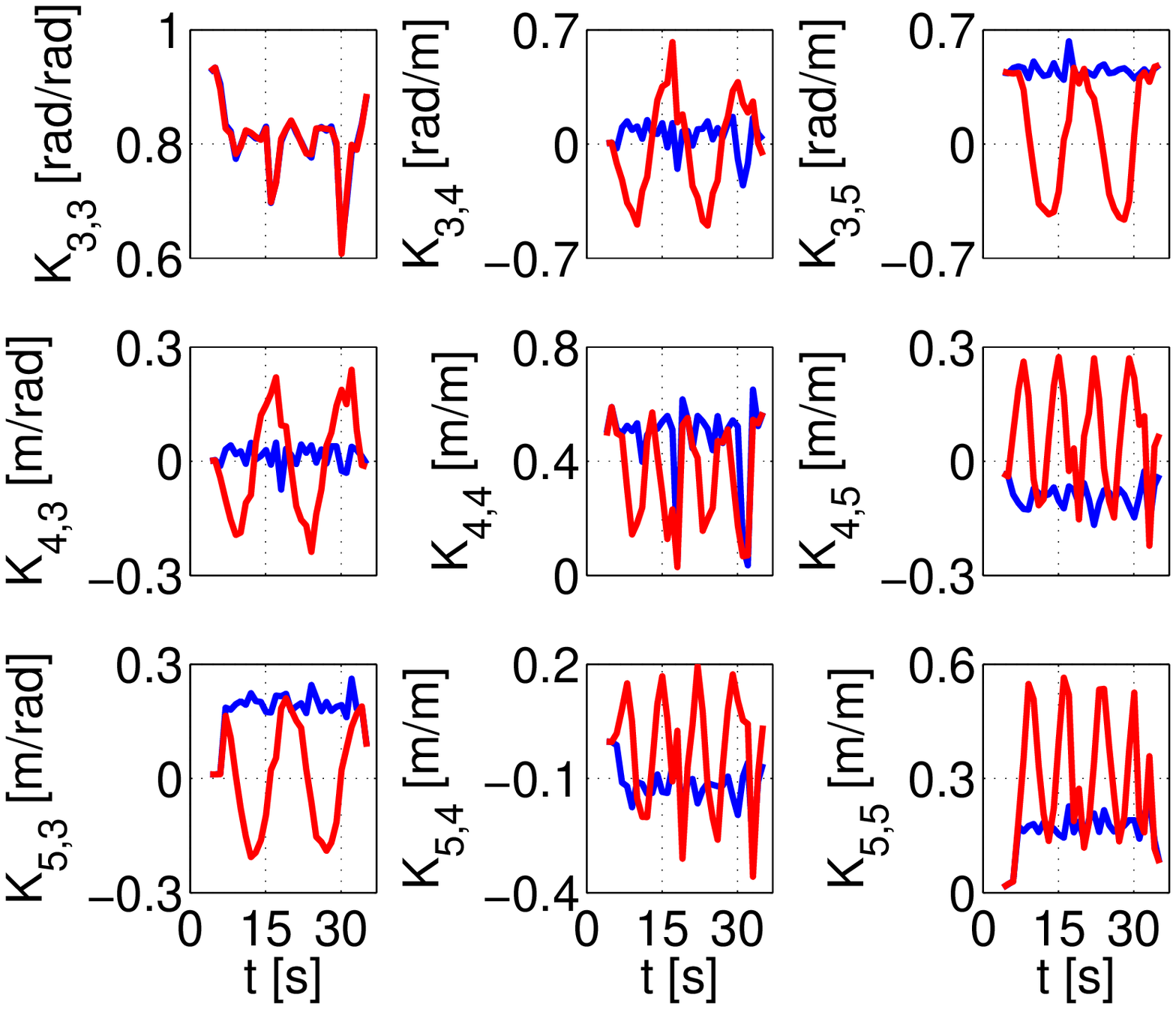}
\end{tabular}
\caption{Circular trajectory experiment: IEKF (\textcolor[rgb]{0,0,1}{\protect\rule[0.25ex]{0.5cm}{2pt}}), MEKF 
(\textcolor[rgb]{1,0,0}{\protect\rule[0.25ex]{0.5cm}{2pt}}), Ground truth (\textcolor[rgb]{0,0,0}{\protect\rule[0.25ex]{0.5cm}{2pt}})}
\label{fig:exp2}
\end{center}
\end{figure}

In the second experiment the robot begins stationary, then executes a circular (thus non-linear) trajectory consisting of two identical circles traveled in the
counter-clockwise direction, obtained by setting differing constant speeds on the left and right wheels. As in the previous experiment, for each of
the two filters we plot the
estimated system states versus the ground truth reported by the OptiTrack system, as well as an overhead view of the positions for visualization purposes and a
$3 \times 3$ subset of the Kalman gain matrix $K$ entries. The results are shown in Figure~\ref{fig:exp2}.

Figure~\ref{fig:exp2} clearly shows that the performance of the two estimators is no longer identical, and that the IEKF-computed estimates are closer to the 
ground truth than the MEKF ones. In the final configuration, the IEKF has an error of $(\Delta x,\Delta y,\Delta \psi)=(14.3\text{ cm}, 1.2\text{ cm}, 
9.7^\circ)$ while the MEKF has $(29.4\text{ cm}, 5.7\text{ cm}, 21.3^\circ)$. For RMS values of the estimation discrepancy vectors, the IEKF shows
$\text{RMS}(\Delta x,\Delta y,\Delta \psi)=(10.6\text{ cm}, 14.2\text{ cm},
5.7^\circ)$ and the MEKF $(18.5\text{ cm}, 21.3\text{ cm},
12.4^\circ)$. Thus in this non-linear trajectory case, the IEKF exhibits an estimation performance advantage over the MEKF.

The Kalman gain matrix entries plotted in Figure~\ref{fig:exp2} demonstrate a feature of the IEKF already discussed in Section~\ref{sec:LIEKF}: the
independence of the linearized Kalman Filter dynamics (as the matrices $(A,B,C,D)$ used to compute $P$ hence $K$) on the estimated system state $\hat{R}$,
c.f.~\eqref{eq:IEKFmatrices} for IEKF versus~\eqref{eq:convEKFerrsys} for the MEKF. Indeed throughout the circular trajectory where $\hat{R}$ varies with time,
the IEKF gains remain approximately level while the MEKF gains oscillate --- note that they can not remain absolutely level, as the IEKF gains depend on the 
scan matching covariance $R_\nu$ which varies throughout the trajectory due to inhomogeneities in the environment. This is logical as the gains should 
definitely adapt to the environment's observability, such that the filter does not trust the scan matching output when the environment is 
underconstrained and contains no information along some specific direction(s). The IEKF
design's greater independence from estimated states is expected to provide better estimation accuracy than in the MEKF design, and this is indeed what we see 
here.

\section{Conclusions}
\label{sec:conclusions}

We have successfully designed and experimentally validated a Kinect depth camera scan matching-aided Invariant EKF-based localization system and compared it 
against a
Multiplicative
EKF-based
design in terms of theoretical features and estimation performance relative to a ground truth. The fundamental advantage of the IEKF is its guaranteed
increase in robustness to poor estimates of state, a fundamental weakness of the
MEKF design. 

We described both filter designs and tested them experimentally, demonstrating that the state estimates follow the ground truth and 
confirming
the consistency of the novel ICP covariance estimation~\eqref{eq:pncovfinal} derived in Section~\ref{sec:ICPcov}. Experimental testing illustrated the
theoretical features and advantages of the IEKF and demonstrated its improved estimation accuracy over the MEKF design for a circular (non-linear)
robot trajectory.

Future work involves mixing the introduced techniques with pose-SLAM and smoothing in order to take advantage of the loop closures which may occur during the 
motion, tackling the case where a 3D map is not available, and applying the IEKF to more
complicated problems such
as outdoor mobile cartography.

\section*{Acknowledgments}

We thank Tony No{\"e}l for his extensive help with setting up the Wifibot platform. The work reported in this paper was
partly supported by the Cap Digital Business Cluster TerraMobilita Project.

\bibliographystyle{IEEEtran}
\bibliography{IEEEabrv,bibliography}

\end{document}